\documentclass[aps,prb,twocolumn,superscriptaddress,floatfix,10pt,letterpaper,showpacs]{revtex4}

\pdfoutput=1

\begin{document}

\begin{titlepage}

\title{Symmetry protected topological orders in interacting bosonic systems}

\author{Xie Chen}
\affiliation{Department of Physics, Massachusetts Institute of
Technology, Cambridge, Massachusetts 02139, USA}

\author{Zheng-Cheng Gu}
\affiliation{Kavli Institute for Theoretical Physics, University of California, Santa Barbara, CA 93106, USA}

\author{Zheng-Xin Liu}
\affiliation{Institute for Advanced Study, Tsinghua University,
Beijing, 100084, P. R. China}
\affiliation{Department of Physics,
Massachusetts Institute of Technology, Cambridge, Massachusetts
02139, USA}

\author{Xiao-Gang Wen}
\affiliation{Department of Physics, Massachusetts Institute of
Technology, Cambridge, Massachusetts 02139, USA}
\affiliation{Perimeter Institute for Theoretical Physics, Waterloo, Ontario, N2L 2Y5 Canada}
\affiliation{Institute for Advanced Study, Tsinghua University,
Beijing, 100084, P. R. China}

\begin{abstract}
Symmetry protected topological (SPT) states are bulk gapped states with gapless
edge excitations protected by certain symmetries. The SPT phases in free
fermion systems, like topological insulators, can be classified by the
K-theory. However, it is not known what SPT phases exist in general interacting systems.
In this paper, we present a systematic way to construct SPT phases
in interacting bosonic systems, which allows us to identify many new SPT
phases, including three bosonic versions of topological insulators in three
dimension and one in two dimension protected by particle number conservation
and time reversal symmetry. Just as group theory allows us to construct 230 crystal structures in 3D, we
find that group cohomology theory allows us to construct different interacting
bosonic SPT phases in any dimensions and for any symmetry groups.  In
particular, we are going to show how topological terms in the path integral
description of the system can be constructed from nontrivial group cohomology
classes, giving rise to exactly soluble Hamiltonians, explicit ground state wave
functions and symmetry protected gapless edge excitations.
\end{abstract}


\maketitle


\end{titlepage}



We used to believe that different phases of matter are different because they
have different symmetries.\cite{L3726,GL5064,LanL58} Recently, we see a deep
connection between quantum phases and quantum
entanglement\cite{KP0604,LW0605,CGW1038} which allows us to go beyond this framework. First it was realized that even
in systems without any symmetry there can be distinct quantum phases -- topological
phases\cite{W9039,WN9077} due to different patterns of long-range
entanglement in the states.\cite{CGW1038}
For systems with symmetries, difference in long-range entanglement and in symmetry still lead to distinct phases. Moreover, even short-range
entangled states with the same symmetry can belong to different phases.
These symmetric short-range entangled states are said to contain a new kind of
order -- symmetry protected topological (SPT) order.\cite{GW0931}
The SPT phases have symmetry protected gapless edge modes despite the
bulk gap, which clearly indicates the topological nature of this order. On the
other hand, the gapless edge modes disappear when the symmetry of the system is
broken, indicating that this is a different type of topological order than
that found in fractional quantum Hall systems\cite{TSG8259,L8395} whose edge
modes cannot be removed with any local perturbation.\cite{W9505} Also, SPT
orders have no factional statistics or fractional charges, while intrinsic
topological orders from long range entanglement can have them. The discovery of 
SPT order hence greatly expands our original 
understanding of possible phases in many-body systems.

One central issue is to understand what SPT phases exist and much progress has
been made in this regard. The first system with SPT order was discovered
decades ago in spin-1 Haldane chains. The Haldane chain with antiferromagnetic
interactions was shown to have a gapped bulk\cite{H8364,AKL8877} and degenerate
modes at the ends of the chain\cite{HKA9081,GGL9114,N9455} which are protected
by spin rotation or time reversal symmetry of the system.\cite{GW0931,PBT1225}
This model has been generalized, leading to a complete classification of SPT
orders in one dimension.\cite{CGW1107,LK1103,TPB1102,SPC1139} Topological
insulator \cite{KM0501,BZ0602,KM0502,MB0706,FKM0703,QHZ0824} with time reversal
symmetry and particle number conservation protected gapless edge modes provides
the first example of SPT order in higher dimensions. The non-interacting
nature of fermions in this system allows a complete classification of such
kind of SPT orders.\cite{K0922,RSF1010}

However, the understanding of SPT orders in free fermion systems is not enough,
because particles in many-body systems do interact and interaction could
dramatically change the phase diagram obtained for free fermion systems.
Examples have been found in one\cite{LK1009} and
two\cite{fSPT,Qi12arXiv,fduality,YR12arXiv,RZ12arXiv} dimensions where
different free fermion SPT phases become the same with interactions. Moreover,
new interacting SPT phases can exist which cannot be realized by free fermion
systems. Therefore, to have a complete understanding of SPT order, we need to
answer the following question: what SPT phases exist in general interacting
systems, with topological features stable under any symmetric interaction as
long as no phase transition occurs?

In this paper, we answer this question by presenting a systematic construction
of SPT phases in interacting bosonic systems in any dimension and for any
symmetry. The stability of the order in the constructed model can be proven
under any type of symmetric interaction (at least in one and two dimension). In
one dimension, our construction reproduces the classification result already
known. While no bosonic SPT phases were known previously in two and higher
dimensions, our construction leads to the discovery of many new SPT phases in
such systems, which are summarized in table I. As listed in the first row, we
find one kind of bosonic topological insulator in 2D and three kinds in 3D with
boson number conservation symmetry $U(1)$ and time reversal symmetry $Z_2^T$.
If boson numbers are allowed to fluctuate but time reversal symmetry $Z_2^T$ is
preserved, then we find one kind of bosonic topological superconductor in every
odd spatial dimension, as listed in the second row.  More generally, for systems
in $d$ spatial dimension and with symmetry of group $G$, we are going to
present a way to write down a quantized topological term in the path integral
in $d+1$ dimensional space-time based on the nontrivial group cohomology of
$G$. From the path integral, we can find the ground state wave function,
identify the gapless edge states and understand how symmetry protects the
gaplessness of the edge and hence the SPT order even against strong
interactions. For simplicity, we are going to first present the construction in
detail for the one dimensional Haldane chain, demonstrate the emergence of its SPT order from
the nontrivial group cohomology of time reversal symmetry and then generalize
to higher dimensions and to all other symmetries.

\begin{table}[tb]
 \centering
 \begin{tabular}{ |c||c|c|c|c| }
 \hline
 Symmetry & $d=0$ & $d=1$ & $d=2$ & $d=3$  \\
\hline
\hline
\color{red}{$U(1)\rtimes Z_2^T$}  & \color{red}{$\Z$} & \color{red}{$\Z_2$} & \color{red}{$\Z_2$} & \color{red}{$\Z^2_2$}   \\
\hline
\color{blue}{$Z_2^T$}  & \color{blue}{$\Z_1$} & \color{blue}{$\Z_2$} & \color{blue}{$\Z_1$} & \color{blue}{$\Z_2$}   \\
\hline
$U(1)$ & $\Z$  & $\Z_1$ & $\Z$ & $\Z_1$   \\
\hline
$SO(3)$ & $\Z_1$  & $\Z_2$ & $\Z$ & $\Z_1$    \\
\hline
$SO(3)\times Z_2^T$ & $\Z_1$  & $\Z_2^2$ & $\Z_2$ & $\Z_2^3$    \\
\hline
$Z_n$ & $\Z_n$  & $\Z_1$ & $\Z_n$ & $\Z_1$    \\
\hline
$Z_2^T\times D_2=D_{2h}$ & $\Z^2_2$  & $\Z^4_2$ & $\Z^6_2$ & $\Z^9_2$    \\
\hline
 \end{tabular}
 \caption{
(Color online)
SPT phases in $d$-spatial dimensions protected by some simple
symmetries (represented by the symmetry groups).  Here $\Z_1$ means that
our construction only gives rise to
the trivial phase.
$\Z_n$ means that the constructed non-trivial SPT phases plus the
trivial phase are labeled by elements in $\Z_n$.  $Z_2^T$ represents time
reversal symmetry,
$U(1)$ represents boson number conservation symmetry, $Z_n$ represents cyclic symmetry, \etc.
The red row is for bosonic topological insulators and the blue row is for bosonic
topological superconductors.
}
\label{tb}
\end{table}



\begin{figure}[htbp]
\begin{center}
\includegraphics[scale=0.3]{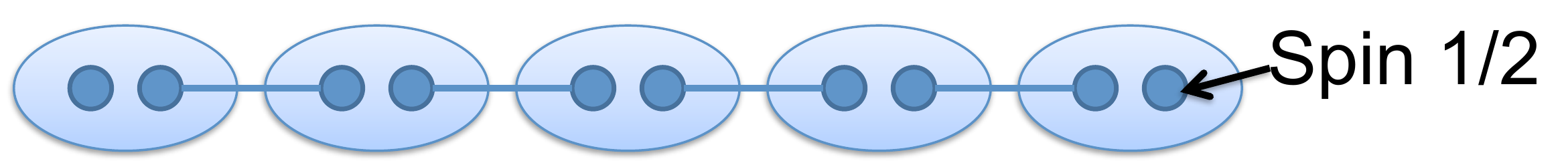}
\end{center}
\caption{
(Color online)Dimer form of the ground state wave-function in Haldane chain. Each site (big oval) contains two spin $1/2$'s (small dot), which are connected into singlet pairs (connected dots) between neighboring sites.
}
\label{dimer}
\end{figure}

The fixed point ground state wave-function of the Haldane chain\cite{GW0931} takes a simple dimer form as shown in Fig.\ref{dimer}, where each site contains two spin $1/2$'s which are connected into singlet pairs $|\uparrow^r_i\downarrow^l_{i+1}\rangle-|\downarrow^r_{i}\uparrow^l_{i+1}\rangle$ between neighboring sites. \footnote{The usual spin $1$ degrees of freedom in the Haldane chain can be obtained by projecting the two spin $1/2$'s on each site to their symmetric subspace, but we will ignore this projection here so that the wave-function is in simpler form and still contains the same topological features.} Time reversal symmetry acts as $M(\cT)=i\sigma_yK$ on each spin $1/2$, where $K$ is complex conjugation and $\sigma_y$ is the $y$ component of the spin operator. The wave-function is invariant under the symmetry action. Note that for each spin $1/2$ $M(\cT)^2=-I$ while on each site with two spins $(M(\cT)\otimes M(\cT))^2=I$. So the states on each site form a representation of $Z_2^T$, the symmetry group generated by time reversal symmetry.

The wave-function on a closed chain is the gapped ground states of anti-ferromagnetic Heisenberg interactions between each pair of spin $1/2$'s,
$H=\sum_i \pmb{\sigma}_i^r \cdot \pmb{\sigma}_{i+1}^l $
where $\pmb{\sigma}_i^l$ and $\pmb{\sigma}_i^r$ are spin operators for the left and right spin $1/2$ on each site respectively. Obviously, the Hamiltonian is time reversal invariant. Therefore we have a situation where the ground state does not break any symmetry of the system, yet the system is far from a trivial phase which becomes evident when we put the system on an open chain. When the chain is open, the dangling spin $1/2$ at each end forms a nontrivial projective representation of $Z_2^T$ with $M(\cT)^2 = -I$, which does not allow a one-dimensional representation (for definition of projective representations, see appendix \ref{prorep}). Therefore, the degeneracy of the edge state is robust under any perturbation as long as time reversal symmetry is preserved.

Indeed, the nontrivial projective representation on the edge is the key to the existence of SPT order in the Haldane chain and applies in general to all one dimensional bosonic SPT phases.\cite{CGW1107,SPC1139} A generic way to write projective representations of group $G$ is to define symmetry operations $M(g)$($g \in G$) on group element labeled states $\ket{g_0}$($g_0 \in G$) as
\be
M(g)\ket{g_0}=\nu_2^{s(g)}(g_0,g^{-1}g^*,g^*)\ket{g g_0}, g^*=E
\label{sym_edge}
\ee
where the $2$-cocycle $\nu_2(g_0,g_1,g_2)$ is a function from three group elements to a $U(1)$ phase factor satisfying 
\begin{align}
\nu_2^{s(g)}(g_0,g_1,g_2)=\nu_2(gg_0,gg_1,gg_2), \ g\in G \label{Gcohsym2} \\
\text{and} \ \ \ \ \frac{\nu_2(g_1,g_2,g_3)\nu_2(g_0,g_1,g_3)}{\nu_2(g_0,g_2,g_3)\nu_2(g_0,g_1,g_2)}=1
\label{Gcoh2}
\end{align}
$s(g)=1$ if $g$ is unitary and $s(g)=-1$ if $g$ is antiunitary. (We prove this in appendix \ref{prorep}.) The projective representation on the edge of the Haldane chain is of exactly this form if we relabel the spin states with group elements. The time reversal symmetry group contains two elements $Z_2^T=\{E,\cT\}$ with $\cT \circ \cT=E$. For the left spin $1/2$ on each site, label $\ket{\up}$/$\ket{\down}$ as $\ket{E}$/$\ket{\cT}$ and for the right one, label $\ket{\up}$/$\ket{\down}$ as $\ket{\cT}$/$-\ket{E}$. The total wave-function becomes
\be
\ket{\Phi_{dimer}}=\prod_i \ket{\cT_i^r\cT_{i+1}^l}+\ket{E_{i}^rE_{i+1}^l}
\label{phi_dimer}
\ee
Time reversal symmetry on the edge spin then acts as $M(\cT)\ket{E}=-\ket{\cT}$ and $M(\cT)\ket{\cT}=\ket{E}$, which indeed takes the form in Eq.(\ref{sym_edge}) with the $2$-cocycle $\nu_2$ of time reversal symmetry given by
\begin{align}
\nu_2(E,\cT,E)=\nu_2(\cT,E,\cT)=-1 \\ \nonumber
\text{all other} \ \nu_2(g_0,g_1,g_2)=1, \ g_i \in Z_2^T
\end{align}


\begin{figure}[htbp]
\begin{center}
\vspace{-0.50in}
\includegraphics[scale=0.28]{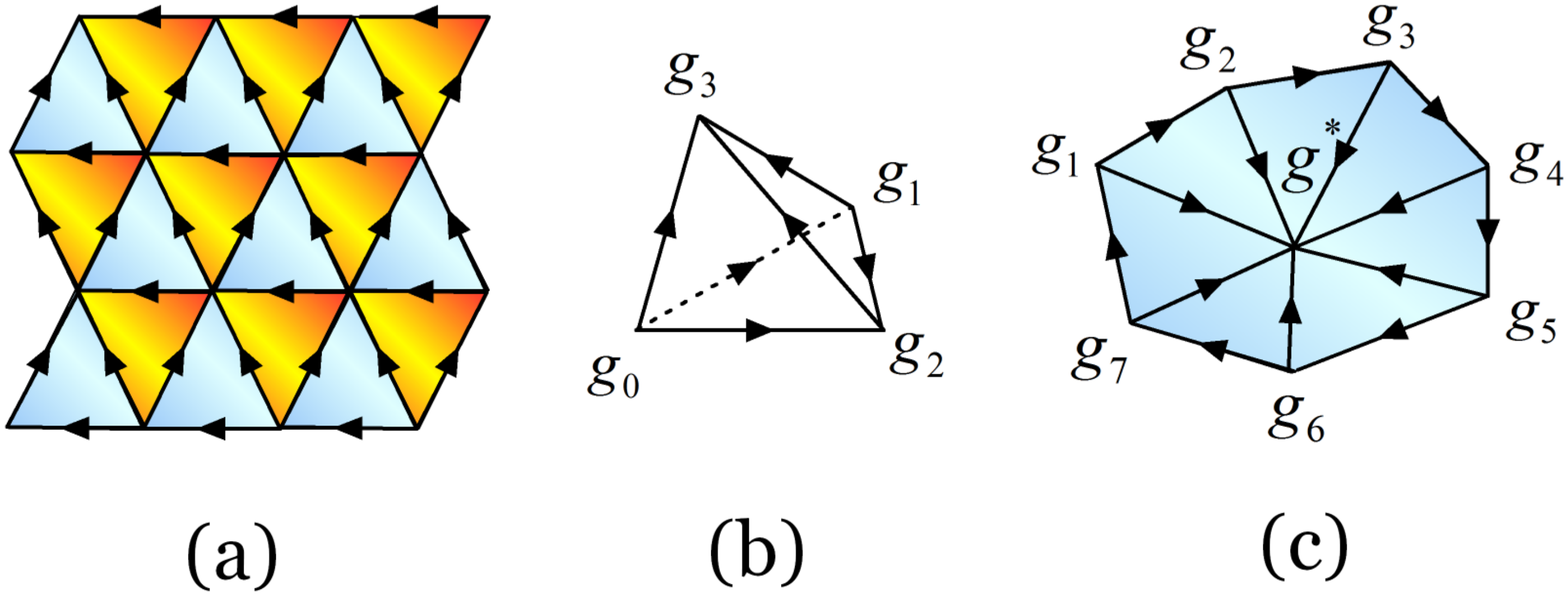}
\vspace{-0.80in}
\end{center}
\caption{
(Color online)
(a) A branched triangularization of space-time. (For details of braching see appendix \ref{fixed_point})
(b) A tetrahedron -- the simplest discrete closed surface.
$\prod \nu^{s_{ijk}}(g_i,g_j,g_k)=1$ on a tetrahedron is guaranteed by Eq.(\ref{Gcoh2}).
Note that $s_{123}=s_{013}=1$ and $s_{023}=s_{012}=-1$.
(c) Discretized space-time manifold $M_\text{ext}$ on an open disk with boundary manifold $M$. $g_i \in M$, $g^*$  is in the interior of $M_\text{ext}$.
}
\label{wavefunction}
\end{figure}

This essential feature of $1$D SPT phases can be reproduced from a simple construction of the low energy effective path integral as we show in the following for the Haldane phase. The path integral formulation not only provides understanding of the nontrivial physics from a bulk perspective but also forms the basis for the generalization of SPT phases into higher dimensions.

To write the path integral for disordered phases, we first realize that the field of the system fluctuates strongly at all length scales and the low energy effective theory has no continuous limit. Therefore, we choose to write the path integral on discrete space-time. Imagine that we discretize the $(1+1)D$ space time of the Haldane chain with a branched triangularization as shown in Fig.\ref{wavefunction}(a). On each vertex of the space-time complex, we put a $g_i \in Z_2^T$. Time reversal acts as complex conjugation $K$ together with a mapping from $g_i$ to $\cT g_i$. The path integral is then written as
\begin{align}
Z &=|G|^{-N_v}\sum_{\{g_i\}} \e^{ - S(\{g_i\}) } \nonumber \\
\e^{ - S(\{g_i\}) }&=\prod_{\{ijk\}} \nu_{2}^{s_{ijk}}(g_i,g_j,g_k)
\label{topo_term2}
\end{align}
$|G|$ is the number of elements in $G$($|G|=2$ for the time reversal symmetry), $N_v$ is the number of
vertices in the complex. Also, $s_{ijk}=\pm 1$ depending on the orientation of the triangle. 
Compared to the continuous formulation of path integral $Z=\int Dg\ \e^{-\int \dd \v x \dd \tau\; \cL[g(\v x,\tau)]}$
$g_i$ corresponds to the field $g(\v x,t)$ and  $\sum_{\{g_i\}}$ corresponds to the
path integral $\int Dg$. $ \e^{ - S(\{g_i\}) }$ is the action-amplitude on the discretized
space-time that corresponds to $\e^{- \int \dd^d \v x \dd \tau\; \cL[g(\v x,\tau)]} $ of the continuous formulation and
$\nu_{2}^{s_{ijk}}(g_i,g_j,g_k)$ corresponds to the action-amplitude
$\e^{- \int_{(i,j,k)} \dd \v x \dd \tau\; \cL[g(\v x,\tau)]}$ on a single
triangle $(i,j,k)$.The above term can be formally regarded as the discrete symmetry group and discrete space time generalization of topological $\theta$-term for an $O(3)$ non-linear sigma model.(For details see appendix \ref{theta}.)

From the properties of $\nu_2(g_0,g_1,g_2)$, it can be checked that the path integral as defined is symmetric under time reversal symmetry with an action amplitude that is in a fixed point form and is always equal to $1$ on a closed surface. First because $\nu_2(\cT g_0,\cT g_1,\cT g_2)=\nu_2^{-1}(g_0,g_1,g_2)$ (see Eq.(\ref{Gcohsym2})), the path integral is invariant under time reversal. Secondly, Eq.(\ref{Gcoh2}) gives rise to a renormalization flow under which the form of the action amplitude remains invariant (for details see appendix \ref{fixed_point}).
Finally, Eq.(\ref{Gcoh2}) also guarantees that the action amplitude on a closed space time surface is always $1$. This can be easily verified on the simplest discrete closed surface -- a tetrahedron, as shown in Fig.\ref{wavefunction}(b). More complicated closed surfaces are obtained by putting tetrahedrons together, hence the action amplitude will always be $1$. Therefore, we have constructed a quantized topological term for the path integral description of $(1+1)D$ systems with time reversal symmetry on discrete space time.


But how do we know that this quantized topological term describes the SPT order in Haldane chain? This can be made explicit by finding the ground state wave-function of the system from the quantized topological term. The ground state wave-function can be obtained by imaginary time evolution from time $-\infty$ until time $0$. In our formulation, this is equivalent to imaginary time path integral on a space-time geometry with a boundary (at time $0$). Denote the boundary as $M$ and the whole manifold (a disk) as $M_\text{ext}$. As we are considering a fixed point path integral, it does not matter how big the interior of $M_\text{ext}$ is and we can reduce it, for example, to just one point as shown in Fig.\ref{wavefunction} (c).

To obtain the ground state wave-function, we fix the degrees of freedom $\{g_i\}_M$ on $M$ and choose an arbitrary configuration, for example $g^*=E$, for degrees of freedom in the interior of $M_\text{ext}$. We find
\begin{align}
\Psi(\{g_i\}_M) 
=\prod_i \nu_2(g_i,g_{i+1},g^*)
\label{PsiM}
\end{align}
where $ \prod_{i}$ is product over all triangles on $M_\text{ext}$ and for simplicity of notation we have chosen all triangles to be oriented clockwise. The wave-function on $M$ does not depend on the choice of $g^*$. Note that time reversal acts as complex conjugate $K$ together with a change of basis $\ket{E} \to \ket{\cT}$, $\ket{\cT} \to \ket{E}$ on each $g_i$ and the wave-function is invariant under this action.

\begin{figure}[htbp]
\begin{center}
\includegraphics[scale=0.35]{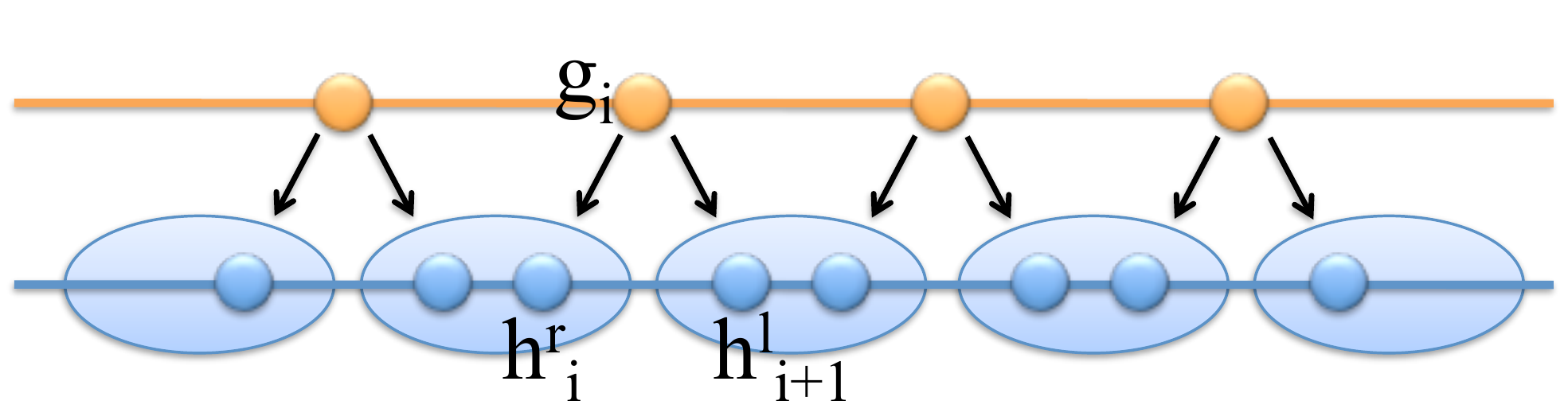}
\end{center}
\caption{(Color online) Duality transformation between wave-functions in Eq.(\ref{phi_dimer}) and Eq.(\ref{PsiM}).} \label{dual_trans}
\end{figure}

This form of wave-function does not look exactly the same as the dimer form shown in Fig.\ref{dimer}, but it will be after a duality transformation as shown in Fig.\ref{dual_trans}. First expand each $g_i$ into two degrees of freedom $h_i^r$ and $h_{i+1}^l$ such that $h_i^r=h_{i+1}^l=g_i$ and the amplitude of each configuration in the wave-function remains unchanged, $\Phi(\{h_i^r=h_{i+1}^l=g_i\})=\prod_{i} \nu_{2}(g_i,g_{i+1},g^*)$. Then combine $h_i^l$ and $h_i^r$ into one site and apply a change of basis on each site
\begin{equation*}
\ket{h_i^l, h_i^r}' = \nu_2(h_i^l,h_i^r,g^*)\ket{h_i^l,h_i^r} = \nu_2(g_{i-1},g_i,g^*)\ket{h_i^l,h_i^r}
\end{equation*}
The amplitude of all configurations in the new basis becomes $1$, $\Phi'(\{h_i^r=h_{i+1}^l=g_i\})=1$ which can be equivalently written as a product of dimers between neighboring sites $\ket{\Phi}'=\prod_i \sum_{g_i}\ket{h_i^r=g_i, h_{i+1}^l=g_i}$. In this way we have mapped each degree of freedom $g_i$ into a dimer and the total wave-function takes the same form as Eq.(\ref{phi_dimer}).

It is also important to check that time reversal symmetry acts as in Eq.(\ref{sym_edge}) on the edge degree of freedom in the ground state. To see this, note that after the duality transformation, time reversal symmetry acts separately on each site and is composed of complex conjugation $K$ together with unitary transformation on the basis states
\begin{eqnarray*}
M_{lr}(g)\ket{h_i^l, h_i^r}' & = & \frac{\nu_2^{s(g)}(h_i^l,h_i^r,g^*)}{\nu_2(gh_i^l,gh_i^r,g^*)} \ket{gh_i^l, gh_i^r}' \\
						    & = & \frac{\nu_2^{s(g)}(h_i^l,g^{-1}g^*,g^*)}{\nu_2^{s(g)}(h_i^r,g^{-1}g^*,g^*)} \ket{gh_i^l, gh_i^r}'
\end{eqnarray*}
which factorizes on the two degrees of freedom and acts in the same way as in Eq.(\ref{sym_edge}) on the edge.
Therefore, our quantized topological term Eq.(\ref{topo_term2}) provides a proper description of the SPT order in the Haldane chain.




The advantage of reformulating the Haldane chain in the path integral language is that it can be easily generalized to all spatial dimensions $d$ and all symmetry groups $G$. The two cocycles $\nu_2(g_0,g_1,g_2)$ used in the construction have higher dimensional analogues -- the $(d+1)$-cocycles $\nu_{d+1}(g_0,...,g_{d+1})$, which are maps from $d+2$ group elements to a $U(1)$ phase factor and satisfy


\begin{eqnarray*}
\nu_{d+1}^{s(g)}(g_0,g_1,...,g_{d+1})= \nu_{d+1}(gg_0,gg_1,...,gg_{d+1}), g\in G \\
\text{and} \ \ \ \ \prod_{i=0}^{d+2} \nu_{d+1}^{(-1)^i}( g_0,.., g_{i-1}, g_{i+1},..., g_{d+2})=1
\end{eqnarray*}

A quantized topological term in the path integral of a $(d+1)D$ system can be written in terms of the $(d+1)$-cocycles by (1) discretizing the $(d+1)D$ space time with triangularization (triangle in $(1+1)D$ and tetrahedron in $(2+1)D$ \etc); (2) putting group element labeled degrees of freedom onto the vertices (3) assigning action amplitude to each simplex with the corresponding cocycle. The path integral then takes the form
\begin{align}
Z =|G|^{-N_v}\sum_{\{g_i\}} \prod_{\{ij...k\}} \nu_{d+1}^{s_{ij...k}}(g_i,g_j,...,g_k)
\label{topo_term}
\end{align}
where $s_{ij...k}=\pm 1$ depending on the orientation of the simplex $ij...k$. Similar to the $(1+1)D$ case, it can be shown that the path integral is symmetric under symmetries in group $G$, the action amplitude $e^{-S(\{g_i\})}=\prod_{\{ij...k\}} \nu_{d+1}^{s_{ij...k}}(g_i,g_j,...,g_k)$ is in a fixed point form and is quantized to $1$ on a closed manifold.

The ground state wave-function can be obtained from the action amplitude on a open geometry as discussed before $\Psi(\{g_i\}_M) =\prod_{\{i...j*\}} \nu_2(g_i,...,g_j,g^*)$, where $\{g_i\}_M$ is on $M$ and $g^*$ is inside $M_\text{ext}$. $\prod_{\{i...j*\}}$ is the product over all simplices.
An exactly soluble Hamiltonian can be constructed to realize this state as the gapped ground state. To see this, note that $\Psi$ can be mapped through a local unitary transformation $U=\prod_{\{i...j*\}} \nu^{-1}_2(g_i,...,g_j,g^*)\ket{g_i...g_j}\bra{g_i...g_j}$ to a total product state $\Psi_0(\{g_i\}_M)=1$. As $\Psi_0$ is the gapped ground state of $H_0=-\sum_k\ket{\phi_k}\bra{\phi_k}$, $\ket{\phi_k}=\sum_{g_i\in G}\ket{g_i}$, $\Psi$ is the gapped ground state of $H=U^{\dagger}H_0U=\sum_k -U^{\dagger}\ket{\phi_k}\bra{\phi_k}U$, which is local and has symmetry $G$.


The nontrivial SPT order of the system can be seen explicitly from its boundary. The path integral of degrees of freedom on the boundary can be obtained by putting the topological term on an open geometry as shown in Fig.\ref{wavefunction} (c) for $(1+1)D$. Note that the manifold $M$ now corresponds to the space time manifold of the boundary degrees of freedom.  The path integral for the boundary then reads
\be
Z_b=|G|^{-N_v^M}\sum_{\{g_i\}_M} \prod_{\{i...j*\}} \nu_{d+1}^{s_{i...j*}}(g_i,..,g_j,g^*)
\ee
where $\{g_i\}_M$ is the field on $M$ and $g^*$ is inside $M_\text{ext}$. The action amplitude for each $\{g_i\}_M$ configuration does not depend on the choice of $g^*$.

This term can be thought of as a discretized version of the Wess-Zumino-Witten(WZW) term\cite{WZ7195,W8322} in nonlinear $\si$-models because 1. it is a path integral of $(d-1)+1$ dimensional systems written on an extended $(d+1)D$ manifold with a boundary 2. the action amplitude does not depend on how the extended field in the interior of the $(d+1)D$ manifold is chosen 3. its field takes value in a group $G$ and the path integral is invariant under the action of $g \in G$. On the other hand, this term is more general than the original continuous WZW term because it applies to discrete groups like $Z_2^T$ while the continuous WZW term only works for continuous groups. We expect that the boundary theory described by such a discretized WZW term will be gapless as long as symmetry is not broken, similar to systems described by continuous WZW terms. Actually this has been firmly established in $(1+1)D$ and $(2+1)D$. In $(1+1)D$, as we have discussed with the example of time reversal symmetry, symmetry action on the edge degree of freedom does not have a one-dimensional representation, therefore the edge state will always be degenerate. In $(2+1)D$, we have proved rigorously using the tool of matrix product unitary operator that the symmetry action on the boundary requires it to be gapless as long as symmetry is not broken.\cite{CLW1141,duality} Therefore, the boundary of the systems we constructed carries gapless excitations protected by certain symmetry, which reflects the nontrivial SPT order of the system.



In conclusion, we have presented a systematic construction of symmetry
protected topological phases in bosonic systems in any spatial dimension $d$
and with any symmetry of group $G$. In particular, we wrote down a quantized
topological term in the path integral of the system based on nontrivial
$(d+1)$-cocycles of the symmetry group. From the quantized topological term we
found that the ground state wave-function is short range entangled and the
boundary of the systems carries gapless excitations protected by the symmetry
of the system, which clearly demonstrates the nontrivial SPT order of the
system. The number of nontrivial cocycles, hence the number of nontrivial SPT
phases, for some simple symmetry groups are summarized in table \ref{tb}. Our
construction is non-perturbative and works for strongly interacting bosonic
systems.

This research is supported by NSF Grant No. DMR-1005541
and NSFC 11074140. Z.C.G. is supported by NSF Grant No.  PHY05-51164.


\appendix

\section{Projective Representation}
\label{prorep}

Matrices $u(g)$ form a projective representation of symmetry group $G$ if
\begin{align}
u(g_1)u(g_2)=\om(g_1,g_2)u(g_1g_2),\ \ \ \ \
g_1,g_2\in G.
\end{align}
Here $\om(g_1,g_2) \in U(1)$ and $\om(g_1,g_2) \neq 1$, which is called the
factor system of the projective representation. The factor system satisfies
\begin{align}
 \om^{s(g_1)}(g_2,g_3)\om(g_1,g_2g_3)&=
 \om(g_1,g_2)\om(g_1g_2,g_3),
 \label{2cocycle_om}
\end{align}
for all $g_1,g_2,g_3\in G$, where $s(g_1)=1$ if $g_1$ is unitary and $s(g_1)=-1$ if $g_1$ is anti-unitary.
If $\om(g_1,g_2)=1, \ \forall g_1,g_2$, this reduces to the usual linear representation of $G$.

A different choice of pre-factor for the representation matrices
$u'(g)= \bt(g) u(g)$ will lead to a different factor system
$\om'(g_1,g_2)$:
\begin{align}
\label{omom}
 \om'(g_1,g_2) =
\frac{\bt(g_1)\bt^{s(g_1)}(g_2)}{\bt(g_1g_2)}
 \om(g_1,g_2).
\end{align}
We regard $u'(g)$ and $u(g)$ that differ only by a pre-factor as equivalent
projective representations and the corresponding factor systems $\om'(g_1,g_2)$
and $\om(g_1,g_2)$ as belonging to the same class $\om$.

Suppose that we have one projective representation $u_1(g)$ with factor system
$\om_1(g_1,g_2)$ of class $\om_1$ and another $u_2(g)$ with factor system
$\om_2(g_1,g_2)$ of class $\om_2$, obviously $u_1(g)\otimes u_2(g)$ is a
projective presentation with factor system $\om_1(g_1,g_2)\om_2(g_1,g_2)$. The
corresponding class $\om$ can be written as a sum $\om_1+\om_2$. Under such an
addition rule, the equivalence classes of factor systems form an Abelian group,
which is called the second cohomology group of $G$ and is denoted as
$\cH^2[G,U(1)]$.  The identity element $1 \in \cH^2[G,U(1)]$ is the class that
corresponds to the linear representation of the group.

$\om(g_1,g_2)$ can be equivalently expressed as $\nu_2(\t g_0, \t g_1, \t g_2)$ with correspondence
\be
\om(g_1,g_2)=\nu_2(E, g_1, g_1g_2)
\ee
where $E$ is the identity element in the symmetry group $G$. $\nu_2(g_0,g_1,g_2)$ satisfies
\begin{align}
\nu_2^{s(g)}(g_0,g_1,g_2)=\nu_2(gg_0,gg_1,gg_2), \ g\in G
\label{Gcohsym2_A}
\end{align}
and the condition Eq.(\ref{2cocycle_om}) satisfied by $\om(g_1,g_2)$ becomes
\be
\frac{\nu_2(g_1,g_2,g_3)\nu_2(g_0,g_1,g_3)}{\nu_2(g_0,g_2,g_3)\nu_2(g_0,g_1,g_2)}=1
\label{2cocycle_nu}
\ee
Similarly, $\bt(g_1)$ can be equivalently written as $\nu_1(E,g_1)$ where
\be
\nu_1^{s(g)}(g_0,g_1)=\nu_1(gg_0,gg_1)
\ee
The equivalence relation Eq.(\ref{omom}) between $\om(g_1,g_2)$'s then becomes
\be
\nu'_2(g_0,g_1,g_2) =
\frac{\nu_1(g_0,g_1)\nu_1(g_1,g_2)}{\nu_1(g_0,g_2)}
 \nu(g_0,g_1,g_2).
\ee

In the following we will show $M(g)$ defined in Eq.(\ref{sym_edge}) forms a projective representation of symmetry group $G$.
It is easy to check that:
\begin{align}
&M(g_1)M(g_2)|g_0\rangle \nonumber\\=&M(g_1)\nu_2^{s(g_2)}(g_0,g_2^{-1}g^*,g^*)\ket{g_2 g_0} \nonumber\\
=&\nu_2^{s(g_1)}(g_2g_0,g_1^{-1}g^*,g^*)\nu_2^{s(g_1g_2)}(g_0,g_2^{-1}g^*,g^*)\ket{g_1g_2 g_0}\nonumber\\
\end{align}
with $g^*=E$. We note that in the last line, the sign factor $s(g_1g_2)$ arises because the complex conjugate $K$ could have a nontrivial
action on $\nu_2^{s(g_2)}(g_0,g_2^{-1}g^*,g^*)$ if $g_1$ is an antiunitary. According to the $2$-cocycle condition Eq. (\ref{2cocycle_nu}), we have:  \begin{align}
&M(g_1)M(g_2)|g_0\rangle \nonumber\\
=&\nu_2(g^*,g_1g^*,g_1g_2g^*)\nu_2(g_1g_2g_0,g^*,g_1g_2g^*)\ket{g_1g_2 g_0}\nonumber\\
=&\nu_2(g^*,g_1g^*,g_1g_2g^*)\nu_2^{s(g_1g_2)}(g_0,(g_1g_2)^{-1}g^*,g^*)\ket{g_1g_2 g_0}\nonumber\\
=&\nu_2(E,g_1,g_1g_2)M(g_1g_2)\ket{g_0}\nonumber\\
=&\omega_2(g_1,g_2)M(g_1g_2)\ket{g_0}
\end{align}
Now it is clear that $M(g)$ defined in Eq.(\ref{sym_edge}) forms a projective representation of the symmetry group $G$. However, $M(g)$ is usually reducible. It can be reduced to a direct sum of several irreducible projective representations which belong to the same class. Only irreducible projective representations describe the edge states of general 1D SPT phases.


\section{Group cohomology }
\label{Gcoh}

The above discussion on the factor system of a projective representation can be
generalized which gives rise to a cohomology theory of groups.  In this section,
we will briefly describe the group cohomology theory.

For a group $G$, let $M$ be a G-module, which is an abelian group (with
multiplication operation) on which $G$ acts compatibly with the multiplication
operation (\ie the abelian group structure) on M:
\begin{align}
\label{gm}
 g\cdot (ab)=(g\cdot a)(g\cdot b),\ \ \ \ g\in G,\ \ \ \ a,b\in M.
\end{align}
For the cases studied in this paper, $M$ is simply the $U(1)$ group and $a$ a
$U(1)$ phase. The multiplication operation $ab$ is the usual multiplication of
the $U(1)$ phases. The group action is trivial $g\cdot a=a$ ($g\in G$, $a\in
U(1)$) if $g$ is unitary and $g\cdot a = a^*$ if $g$ is anti-unitary.

Let $\nu_n(g_0,...,g_n)$ be a function of $(n+1)$ group
elements whose value is in the G-module $U(1)$. In other words, $\nu_n:
G^{(n+1)}\to U(1)$. $\nu_n$ satisfies
\be
\nu_{n}^{s(g)}(g_0,g_1,...,g_{n})= \nu_{n}(gg_0,gg_1,...,gg_{n}), g\in G
\ee
We will call such a map $\nu_n$ an $n$-cochain:
Let $\cC^n(G,U(1))=\{\nu_n\}$ be the space of all $n$-cochains. Note that $\cC^n(G,U(1))$ is an Abelian group
under the function multiplication
$ \nu''_n(g_0,...,g_n)= \nu_n(g_0,...,g_n) \nu'_n(g_0,...,g_n) $.
We define a map $d_n$ from $\cC^n[G,U(1)]$ to $\cC^{n+1}[G,U(1)]$:
\begin{align}
\label{dnnun}
&\ \ \ \ (d_n \nu_n) ( g_0, g_1,..., g_{n+1})
\nonumber\\
& = \prod_{i=0}^{n+1}
\nu_n^{(-1)^i}( g_0,.., g_{i-1}, g_{i+1},..., g_{n+1})
\end{align}

Let
\begin{align*}
 \cB^n(G,U(1))=\{ \nu_n| \nu_n=d_{n-1} \nu_{n-1}, \nu_{n-1} \in \cC^{n-1}(G,U(1)) \}
\end{align*}
and
\begin{align*}
 \cZ^n(G,U(1))=\{ \nu_{n}|d_n \nu_n=1,  \nu_{n} \in \cC^{n}(G,U(1)) \}
\end{align*}
$\cB^n(G,U(1))$ and $\cZ^n(G,U(1))$ are also Abelian groups
which satisfy $\cB^n(G,U(1)) \subset \cZ^n(G,U(1))$ where
$\cB^1(G,U(1))\equiv \{ 1\}$.
The $n$-cocycle of $G$ is defined as
\begin{align}
 \cH^n(G,U(1))= \cZ^n(G,U(1)) /\cB^n(G,U(1))
\end{align}

Let us discuss some simple cases.
From
\begin{align*}
 (d_1 \nu_1)(g_0,g_1,g_2)= \nu_1(g_0,g_1)\nu_1(g_1,g_2)/\nu_1(g_0,g_2)
\end{align*}
we see that
\begin{align*}
 \cZ^1(G,U(1))=\{  \nu_1| \nu_1(g_0,g_1)\nu_1(g_1,g_2)=\nu_1(g_0,g_2) \} .
\end{align*}
Since $\cB^1(G,U(1))\equiv \{ 1\}$ is trivial,
$\cH^1(G,U(1))=\cZ^1(G,U(1))$. If we define $\alpha(g)=\nu_1(E,g)$, it is easy to
see that the first-cocycle is related to the one-dimensional representations of the group $G$.

From
\begin{align*}
&\ \ \ \ (d_2 \nu_2)(g_0,g_1,g_2,g_3)
\nonumber\\
&=
\nu_2(g_1,g_2,g_3) \nu_2(g_0,g_1,g_3)/\nu_2(g_0,g_2,g_3)\nu_2(g_0,g_1,g_2)
\end{align*}
we see that
\begin{align*}
& \cZ^2(G,U(1))=\{  \nu_2|
\\
&\ \nu_2(g_1,g_2,g_3) \nu_2(g_0,g_1,g_3)=\nu_2(g_0,g_2,g_3)\nu_2(g_0,g_1,g_2)
 \} .
\nonumber
\end{align*}
and
\begin{align*}
& \cB^2(G,U(1))=\{ \nu_2| \\
& \nu_2(g_0,g_1,g_2)=\nu_1(g_0,g_1)\nu_1(g_1,g_2)/\nu_1(g_0,g_2)
 \} .
\end{align*}
The 2-cocycles
$\cH^2(G,U(1))=\cZ^2(G,U(1))/\cB^2(G,U(1))$ classify the
projective representations discussed in section \ref{prorep}.

From
\begin{align}
&\ \ \ \ (d_3 \nu_3)(g_0,g_1,g_2,g_3,g_4)
\nonumber\\
&= \frac{ \nu_3(g_1,g_2,g_3,g_4) \nu_3(g_0,g_1,g_3,g_4)\nu_3(g_0,g_1,g_2,g_3) }
{\nu_3(g_0,g_2,g_3,g_4)\nu_3(g_0,g_1,g_2,g_4)}
\end{align}
we see that
\begin{align*}
& \cZ^3(G,U(1))=\{  \nu_3|
\\
&\ \frac{ \nu_3(g_1,g_2,g_3,g_4) \nu_3(g_0,g_1,g_3,g_4)\nu_3(g_0,g_1,g_2,g_3) }
{\nu_3(g_0,g_2,g_3,g_4)\nu_3(g_0,g_1,g_2,g_4)}
=1
 \} .
\end{align*}
and
\begin{align*}
& \cB^3(G,U(1))=\{ \nu_3| \\
& \nu_3(g_0,g_1,g_2,g_3)=\frac{
\nu_2(g_1,g_2,g_3) \nu_2(g_0,g_1,g_3)}{\nu_2(g_0,g_2,g_3)\nu_2(g_0,g_1,g_2)}
 \}
\end{align*}
which give us the 3-cocycle
$\cH^3(G,U(1))=\cZ^3(G,U(1))/\cB^3(G,U(1))$.


\section{Relationship with the topological-$\theta$ term of the $O(3)$ non-linear sigma model}
\label{theta}
In the section, we will show the amplitudes Eq.(\ref{topo_term2}) can be formally regarded as the discrete group and discrete space time generalization of topological
$\theta$ term of non-linear sigma model. We will first review that the Haldane phase (a non-trivial 1D SPT
phase) is described by a $2\pi$-quantized topological term in continuous
non-linear $\si$-model.\cite{Ng9455} However, such kind of $2\pi$-quantized
topological terms cannot describe more general 1D SPT phases.  We argue that to
describe SPT phases correctly, we must generalize the $2\pi$-quantized
topological terms to discrete space-time.

\begin{figure}[htbp]
\begin{center}
\includegraphics[scale=0.16]{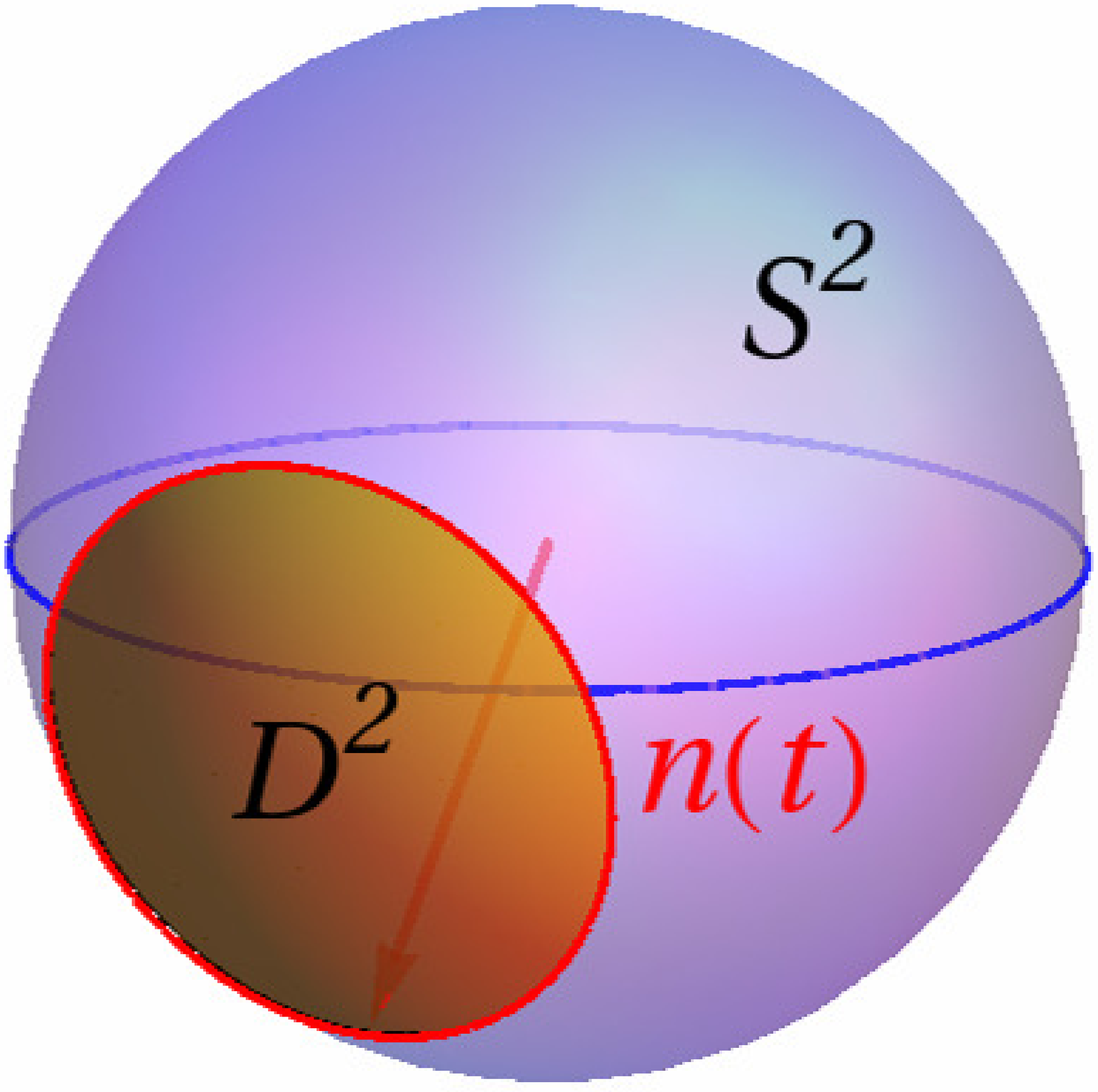}
\end{center}
\caption{
If we extend $\v n(t)$ that traces out a loop
to $\v n(t,\xi)$ that covers the shaded disk,
then the WZW term $\int_{D^2} \dd t\dd \xi\; \v n(t,\xi)\cdot
[\prt_t \v n(t,\xi) \times  \prt_\xi \v n(t,\xi)]$ corresponds to the area of
the disk.
}
\label{topterm1}
\end{figure}

Before considering a spin-$1$ chain, let us first consider a $(0+1)$D non-linear
$\si$-model that describes a single spin, whose imaginary-time action is given
by
\begin{align}
\label{spins}
S &= \oint \dd t  \frac{1}{2g}(\prt_t \v n(t))^2
\nonumber\\
&\ \ \ \ \   + \imth s \int_{D^2} \dd t\dd \xi\; \v n(t,\xi)\cdot
[\prt_t \v n(t,\xi) \times  \prt_\xi \v n(t,\xi)]
\end{align}
where $\v n(t)$ is an unit 3d vector and we have assumed that the time
direction form a circle.  The second term is the Wess-Zumino-Witten (WZW)
term.\cite{WZ7195,W8322} We note that the WZW term cannot be calculated from
the field $\v n(t)$ on the time-circle.  We have to extend $\v n(t)$ to a disk
$D^2$ bounded by the time-circle: $\v n(t) \to \v n(t,\xi)$ (see Fig.
\ref{topterm1}).  Then the WZW term can be calculated from $\v n(t,\xi)$. When
$2s$ is an integer, WZW terms from different extensions only differ by a
multiple of $2\imth \pi$.  So $\e^{-S}$ is determined by $\v n(t)$ and is
independent of how we extend $\v n(t)$ to the disk $D^2$.

The ground states of the above non-linear $\si$-model have $2s+1$ fold
degeneracy, which form the spin-$s$ representation of $SO(3)$.  The energy gap
above the ground state approaches to infinite as $g\to \infty$. Thus a pure WZW
term describes a pure spin-$s$ spin.

\begin{figure}[htbp]
\begin{center}
\includegraphics[scale=0.45]{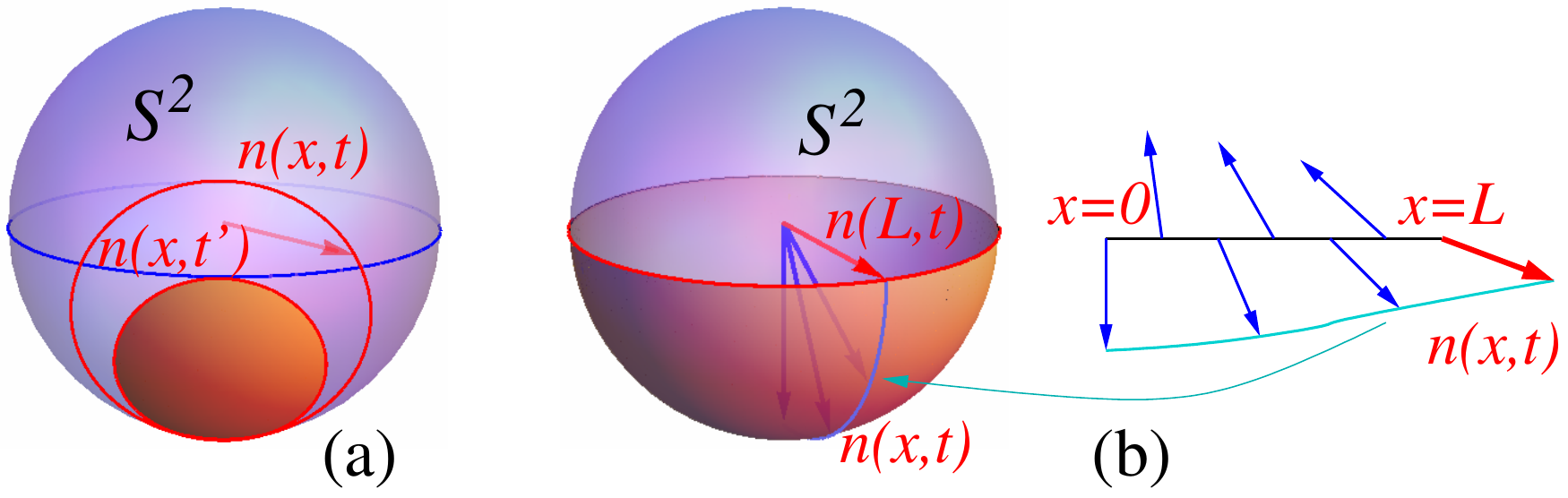}
\end{center}
\caption{
(a) The topological term $W$ describes the number of times that $\v n(x,t)$
wraps around the sphere (as we change $t$).  (b) On an open chain $x \in
[0,L]$, the  topological term $W$ in the (1+1)D bulk becomes the WZW term for
the end spin $ \v n_L(t)=\v n(L,t)$ (where the end spin at $x=0$ is hold
fixed).
}
\label{toptermChain}
\end{figure}

To obtain the action for the $SO(3)$ symmetric antiferromagnetic spin-1 chain,
we can assume that the spins $\v S_i$
are described
by a smooth unit vector field $\v n(x,t)$:
$\v S_i = (-)^i \v n(i a,t)$ (see Fig. \ref{toptermChain}b).
Putting the above single-spin
action for different spins together, we obtain the following (1+1)D non-linear
$\si$-model\cite{Haldane8353}
\begin{align}
\label{2DS}
S=\int \dd x\dd t \frac{1}{2g}(\prt \v n(x,t))^2 +  \imth\th W ,\ \ \
\th=2\pi,
\end{align}
where $W=(4\pi)^{-1} \int \dd t\dd x\; \v n(t,x)\cdot [\prt_t \v n(t,x) \times
\prt_x \v n(t,x)]$ and $ \imth \th  W$ is the topological term.\cite{Haldane8353}  If
the space-time manifold has no boundary, then $\e^{-\imth \th W}=1$ when $\th=0
\text{ mod } 2\pi$.  We will call such a topological term -- a $2\pi$-quantized
topological term.
The above non-linear $\si$-model describes a gapped phase
with short range correlation and the $SO(3)$ symmetry, which is the Haldane
phase.\cite{H8364} In the low energy limit, $g$ flows to infinity and the
fixed-point action contains only the $2\pi$-quantized topological term.  Such a
non-linear $\si$-model will be called topological non-linear $\si$-model.

It appears that the $2\pi$-quantized topological term has no contribution
to the path integral and can be dropped.  In fact, the $2\pi$-quantized topological
term has physical effects and cannot be dropped.  On an open chain, the
$2\pi$-quantized topological term $2\pi \imth W$ becomes a WZW term
for the boundary spin $\v n_L(t) \equiv \v n(x=L,t)$ (see Fig.
\ref{toptermChain}).\cite{Ng9455} The motion of $\v n_L$ is described by
Eq.(\ref{spins}) with $s=1/2$.  So the Haldane phase of spin-1 chain has a spin-1/2
boundary spin at each chain end!\cite{HKA9081,GGL9114,Ng9455}

We see that the Haldane phase is described by a fixed point action which is a
topological non-linear $\si$-model containing only the $2\pi$-quantized
topological term.  \emph{The non-trivialness of the Haldane phase is encoded
in the non trivially quantized topological term}.\cite{Ng9455,SRF0825}
From this example,
one might guess that various SPT phases can be
classified by various topological non-linear $\si$-models, and thus by various
$2\pi$-quantized topological terms.
But such a guess is not correct.

This is because the fixed-point action (the topological non-linear $\si$-model)
describes a short-range-correlated state.
Since the renormalized cut-off length scale of the fixed-point action is always
larger than the correlation length,
the field $\v n( x,t)$ fluctuates strongly even at the
cut-off length scale.  Thus, the fixed-point action has no continuum limit, and
must be defined on discrete space-time.  On the other hand, in our fixed-point
action, $\v n(x,t)$ is assumed to be a continuous field in space-time. The very
existence of the continuum $2\pi$-quantized topological term depends on the
non-trivial mapping classes from the \emph{continuous} space-time manifold
$T^2$ to the \emph{continuous} target space $S^2$.  It is not self consistent
to use such a continuum topological term to describe the fixed-point action for
the
Haldane phase.

As a result, the continuum $2\pi$-quantized topological terms fail to classify
bosonic SPT phases. For example, different possible continuum $2\pi$-quantized
topological terms in Eq.(\ref{2DS}) are labeled by integers, while the integer spin
chain has only two gapped phases protected by spin rotation symmetry: all
even-integer topological terms give rise to the trivial phase and all
odd-integer topological terms give rise to the Haldane phase.  Also non-trivial
SPT phases may exist even when there is no continuum $2\pi$-quantized
topological terms (such as when the symmetry $G$ is discrete).

However, the general idea of using fixed-point actions
to classify
SPT phases is still correct.  But, to use $2\pi$-quantized topological terms
to describe bosonic SPT phases, we need to generalize  them to discrete
space-time.  In the following, we will show that this indeed can be done, using
the (1+1)D model \eq{2DS} as an example.

A discrete (1+1)D space-time is given by a branched triangularization
(see Fig.  \ref{wavefunction} (a)).  Since $S=\int \dd x\dd t L$, on triangularized
space-time, we can rewrite
\begin{align}
& e^{-S}=\prod \nu^{s(i,j,k)}( \v n_i, \v n_j, \v n_k),
\nonumber\\
& \nu^{s(i,j,k)}( \v n_i, \v n_j, \v n_k) =e^{-\int_\triangle \dd x\dd t\; L}
\in U(1),
\end{align}
where $\int_\triangle \dd x\dd t\; L$ is the action on a single triangle.
We see that, on discrete space-time, the action and the path integral are
described by a 3-variable function $\nu( \v n_i, \v n_j, \v n_k)$,
which is called action amplitude.
The $SO(3)$ symmetry requires that
\begin{align}
\label{2cch}
 \nu( g\v n_i, g\v n_j, g\v n_k)=\nu( \v n_i, \v n_j, \v n_k),
\ \ \ g\in SO(3).
\end{align}
In order to use the action amplitude $\nu^{s(i,j,k)}( \v n_i, \v n_j, \v n_k) $
to describe  a $2\pi$-quantized topological term, we must have $\prod
\nu^{s(i,j,k)}( \v n_i, \v n_j, \v n_k)=1$ on any sphere.  This can be
satisfied iff $\prod \nu^{s(i,j,k)}( \v n_i, \v n_j, \v n_k)=1$ on a
tetrahedron -- the simplest discrete sphere (See Fig. \ref{wavefunction}(b)):
\begin{align}
\label{2ccy}
\frac{\nu( \v n_1, \v n_2, \v n_3) \nu( \v n_0, \v n_1, \v n_3)}
{\nu( \v n_0,  \v n_2, \v n_3) \nu( \v n_0, \v n_1, \v n_2 )}
 =1.
\end{align}
(Another way to define topological term on discretized space-time can be
found in \Ref{S8437}.)

A $\nu( \v n_0,  \v n_1, \v n_2)$ that satisfies Eq.(\ref{2cch}) and Eq.(\ref{2ccy}) is
called a 2-cocycle.  If $\nu( \v n_0,  \v n_1, \v n_2)$ is a 2-cocycle, then
\begin{align}
 \nu'( \v n_0,  \v n_1, \v n_2)=
 \nu( \v n_0,  \v n_1, \v n_2)
\frac{\mu(\v n_1,\v n_2) \mu(\v n_0,\v n_1)}{\mu(\v n_0,\v n_2)}
\end{align}
is also a 2-cocycle, for any $\mu(\v n_0,\v n_1)$ satisfying $\mu(g\v n_0,g\v
n_1)=\mu(\v n_0,\v n_1)$, $g\in SO(3)$.  Since $\nu( \v n_0,  \v n_2, \v n_3)$
and $\nu'( \v n_0,  \v n_2, \v n_3)$ can continuously deform into each other,
they correspond to the same kind of $2\pi$-quantized topological term. So we
say that $\nu( \v n_0,  \v n_2, \v n_3)$ and $\nu'( \v n_0,  \v n_2, \v n_3)$
are equivalent.  The equivalent classes of the 2-cocycles $\nu( \v n_0,  \v
n_2, \v n_3)$ give us $\cH^2[S^2,U(1)]$ -- the 2-cohomology group of sphere
$S^2$ with $U(1)$ coefficient.  $\cH^2(S^2,U(1))$ classifies the
$2\pi$-quantized topological term in \emph{discrete} space-time and with $S^2$
as the target space.

Does $\cH^2[S^2,U(1)]$ classify the SPT phases with $SO(3)$ symmetry?  The
answer is no. We know that $S^2$ is just one of the symmetric spaces of
$SO(3)$. To classify the SPT phases, we need to replace the target space $S^2$
by the maximal symmetric space, which is the group itself $SO(3)$ (see
\Ref{CGLW1172} for more discussions).  So we need to consider discrete non-linear
$\si$-model described by $\nu(g_i,g_j,g_k)$, $g_i,g_j,g_k\in SO(3)$.  Now the
2-cocycle conditions becomes
\begin{align}
&
 \nu( gg_i, gg_j, gg_k)=\nu( g_i, g_j, g_k) \in U(1),
\nonumber\\
& \frac{\nu( g_1, g_2, g_3) \nu( g_0, g_1, g_3)}
{\nu( g_0,  g_2, g_3) \nu( g_0, g_1, g_2 )}
 =1,
\end{align}
which defines a ``group cohomology'' $\cH^2[SO(3),U(1)]$.  It
classifies the $2\pi$-quantized topological term for the maximal symmetric
space.  It also classifies the SPT phases with $SO(3)$ symmetry in (1+1)D.


\section{Branched triangulation and topological invariant amplitudes}
\label{fixed_point}

As we have shown, the bosonic symmetry protected order in $d+1$ D can be
described by the amplitude:
\begin{align}
Z=\frac{\sum_{\{g_i\}}}{|G|^{N_v}}\prod \nu_{d+1}^{s_{ij\ldots
k}}(g_i,g_j,\cdots,g_k),\label{action}
\end{align}
where $g_i,g_j,\cdots,g_k\in G$ are group elements of the symmetry
group $G$ and $|G|$ is the order of $G$. The $d+1$-cocycle
$\nu_{d+1}^{s_{ij\ldots k}}(g_i,g_j,\cdots,g_k)$ is defined on the
branched $d+1$-simplex, with $s_{ij\ldots k}=\pm 1$ uniquely
determined by the orientation of the corresponding $d+1$-simplex. In the following, we
will show they are topological invariant amplitudes. Hence, they represent a class of
fixed point amplitudes of (symmetry protected) topologically ordered phases.

\begin{figure}[htbp]
\begin{center}
\includegraphics[scale=0.75]{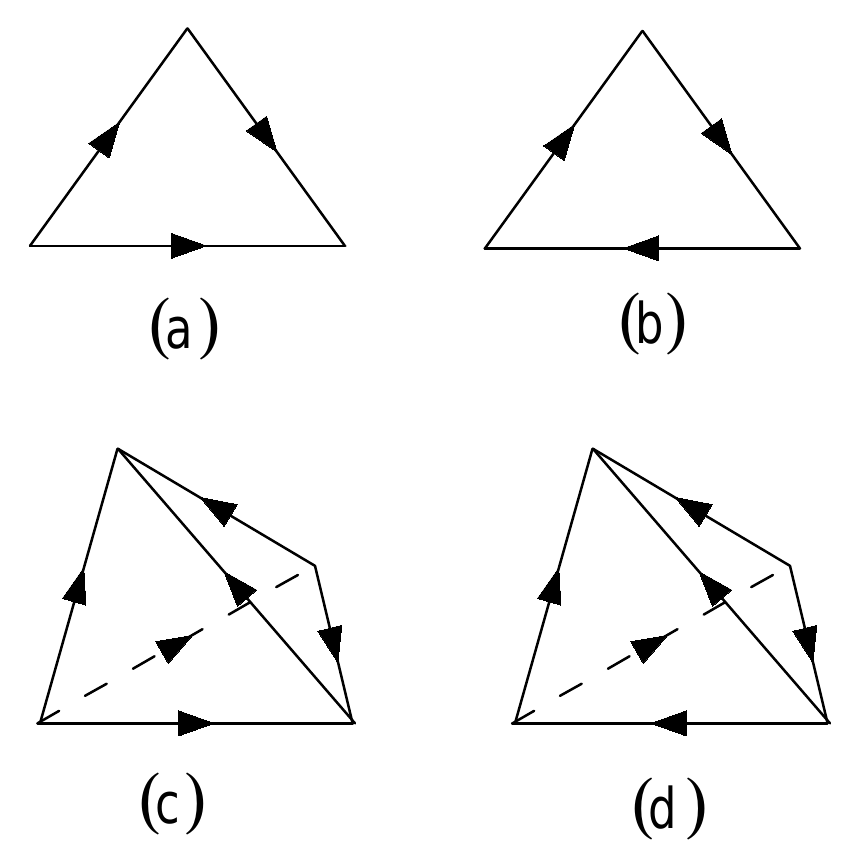}
\end{center}
\caption{Examples of allowed((a),(c)) and unallowed((b),(d))
branching for $2$-simplex and $3$-simplex. \label{branch}}
\end{figure}

\begin{figure}[htbp]
\begin{center}
\includegraphics[scale=0.5]{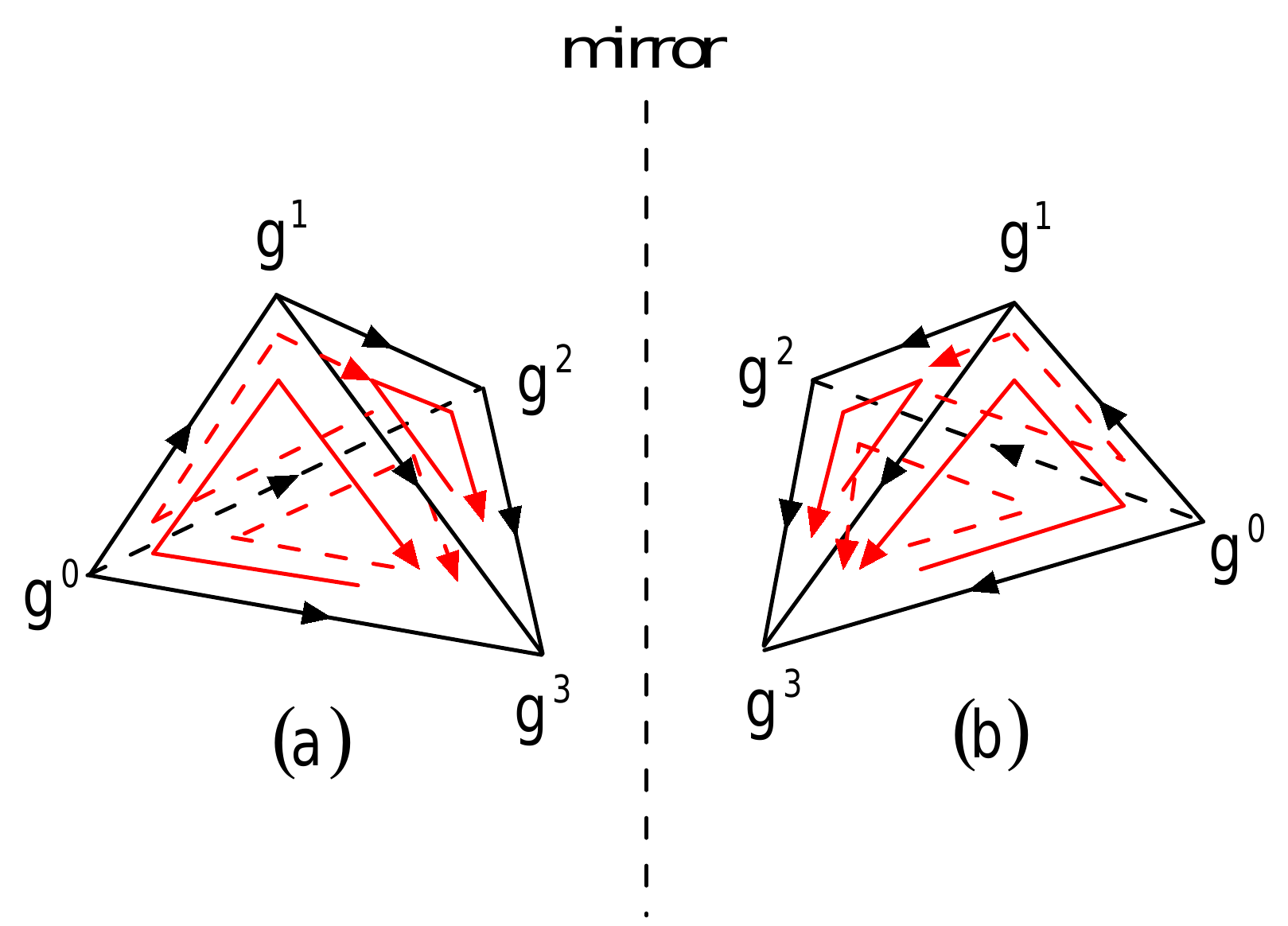}
\end{center}
\caption{(Color online)(a): A branching structure of a $n$-simplex will induce a
canonical ordering for its vertices. For example, if a $3$-simplex
contains a vertex with no incoming edge, then we can label this
vertex as $v^0$ and canonically label the vertices of the rest
$2$-simplex as $v^1,v^2,v^3$. Such a scheme can be applied for
arbitrary $n$-simplex if $n$-simplex has a canonical label. (b): A
branching structure will also induce a canonical orientation for the
$n$-simplex. For example, we can use the right hand rule to
determine the orientations of surfaces of the tetrahedron, then the
orientation of the tetrahedron can be uniquely determined by the
surface opposite to $g^0$\label{ordering}}
\end{figure}

First, we need to give a branching structure to the discretized space-time.
A branching is a choice of an orientation of each edge of an
$n$-simplex such that there is no oriented loop on any triangle. For
example, Fig. \ref{branch} (a) is a branched $2$-simplex and (c) is
a branched $3$-simplex. However, (b) is not an allowed branching
because all its three edges contain the same orientations and thus
form an oriented loop. (d) is also not allowed because one of its
triangle contains an oriented loop. It is easy to check that any
consistent branched structure can induce a canonical ordering for
the vertices of $n$-simplex. Indeed, the branching structure will
also induce an canonical orientation of the $n$-simplex, see Fig.
\ref{ordering}.

\begin{figure}[htbp]
\begin{center}
\includegraphics[scale=0.6]{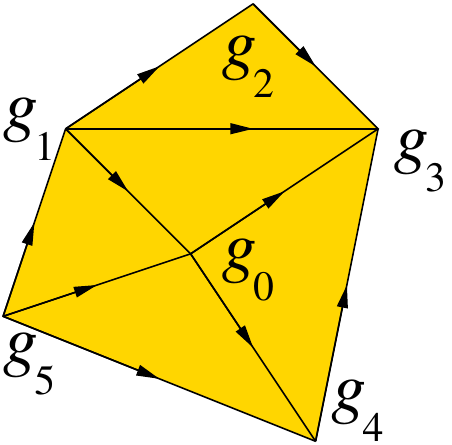}
\end{center}
\caption{ (Color online) The graphic representation of the action-amplitude
$\e^{-S (\{g_i\}) }$ on a complex with a branching structure represented by the
arrows on the edge.  The vertices of the complex are labeled by $i$.  Note that
the arrows never for a loop on any triangle.
} \label{Lg}
\end{figure}

To understand the geometric meaning of the topological invariance of the above amplitude, let us first discuss (1+1)D fixed-point action-amplitude with a symmetry group
$G$. For a $(1+1)$D system on a complex with a branching structure, a
fixed-point action-amplitude (see Fig.  \ref{Lg})
\begin{align}
\label{Lnu2}
&\ \ \ \ \e^{-S(\{g_i\}) }=\prod_{\{ijk\}} \nu_2^{s_{ijk}}(g_i,g_j,g_k)
\nonumber\\
&=
\nu_2^{-1} (g_1,g_2,g_3)
\nu_2 (g_0,g_4,g_3)
\nu_2^{-1} (g_5,g_1,g_0) \times
\nonumber\\
&\ \ \ \ \ \ \ \
\nu_2 (g_1,g_0,g_3)
\nu_2^{-1} (g_5,g_0,g_4)
\end{align}
where each triangle contribute to a phase factor
$\nu_2^{s_{ijk}}(g_i,g_j,g_k)$, $\prod_{\{ijk\}}$ multiply over all the
triangles in the complex Fig.  \ref{Lg}.  Note that the first variable $g_i$ in
$\nu_2(g_i,g_j,g_k)$ correspond to the vertex with two out going edges, the
second variable $g_j$ to the vertex with one out going edges, and the third
variable $g_k$ to the vertex with no out going edges.  $s_{ijk}=\pm 1$
depending on the orientation of $i\to j \to k$ to be anti-clock-wise or
clock-wise.

In order for the action-amplitude to represent a quantized topological
$\th$-term, we must choose $\nu_2(g_i,g_j,g_k)$ such that
\begin{align}
\e^{-S(\{g_i\}) }=\prod_{\{ijk\}} \nu_2^{s_{ijk}}(g_i,g_j,g_k)=1
\end{align}
on closed space-time complex without boundary, in particular, on a tetrahedron
with four triangles (see Fig. \ref{wavefunction}(b)):
\begin{align}
\label{condition}
\e^{-S(\{g_i\}) } &=\prod_{\{ijk\}} \nu_2^{s_{ijk}}(g_i,g_j,g_k)
\nonumber\\
&=
\frac{
\nu_2(g_1,g_2,g_3) \nu_2(g_0,g_1,g_3)
}{
\nu_2(g_0,g_1,g_2) \nu_2(g_0,g_2,g_3)
}=1
\end{align}
Also, in order for our system to have the symmetry generated by the
group $G$, its action-amplitude must satisfy
\begin{align}
\label{nu2a}
 \e^{-S(\{g_i\}) } &= \e^{-S(\{gg_i\}) },
 \text{ if } g \text{ contains no T}
\nonumber\\
 \Big(\e^{-S(\{g_i\}) }\Big)^\dag &= \e^{-S(\{gg_i\}) },
 \text{ if } g \text{ contains one T}
\end{align}
where $T$ is the time-reversal transformation.
This requires
\begin{align}
\label{nu2b}
\nu_2^{s(g)}(g_i,g_j,g_k)=\nu_2(gg_i,gg_j,gg_k).
\end{align}
Eq.(\ref{nu2a}) and Eq.(\ref{nu2b}) happen to be the conditions of
2-cocycles $\nu_2(g_0,g_1,g_2)$ of $G$.
Thus the action-amplitude Eq.(\ref{Lnu2}) constructed from a 2-cocycle
$\nu_2(g_0,g_1,g_2)$ is a quantized topological $\th$-term.

If $\nu_2( g_0,  g_2, g_3)$ satisfy Eq.(\ref{nu2a}) and Eq.(\ref{nu2b}), then
\begin{align}
 \nu_2'( g_0,  g_2, g_3)=
 \nu_2( g_0,  g_2, g_3)
\frac{\mu_1(g_1,g_2) \mu_1(g_0,g_1)}{\mu_1(g_0,g_2)}
\end{align}
also satisfy Eq.(\ref{nu2a}) and Eq.(\ref{nu2b}), for any $\mu_1(g_0,g_1)$ satisfying
$\mu_1(gg_0,gg_1)=\mu_1(g_0,g_1)$, $g\in G$.  So $\nu_2'( g_0,  g_2, g_3)$ also
gives rise to a quantized topological $\th$-term.  As we continuously deform
$\mu_1(g_0,g_1)$, the two quantized topological $\th$-terms can be smoothly
connected.  So we say that the two quantized topological $\th$-terms obtained
from $\nu_2( g_0,  g_2, g_3)$ and  $\nu_2'( g_0,  g_2, g_3)$ are equivalent.
We note that $\nu_2( g_0,  g_2, g_3)$ and  $\nu_2'( g_0,  g_2, g_3)$ differ by
a 2-coboundary $\frac{\mu_1(g_1,g_2) \mu_1(g_0,g_1)}{\mu_1(g_0,g_2)}$.  So the
equivalence classes of $\nu_2( g_0,  g_2, g_3)$ is nothing but the cohomology
group $\cH^2[G,U_T(1)]$.  Therefore, the quantized topological $\th$-terms are
classified by $\cH^2[G,U_T(1)]$.

\begin{figure}[htbp]
\begin{center}
\includegraphics[scale=0.7]{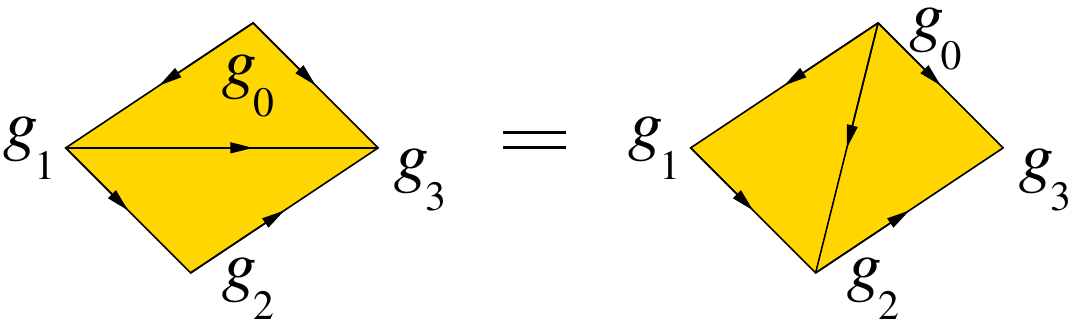}
\end{center}
\caption{
(Color online)
Graphic representation of
 $ \nu_2(g_0,g_1,g_2) \nu_2(g_0,g_2,g_3) = \nu_2(g_1,g_2,g_3)
\nu_2(g_0,g_1,g_3)$
The arrows on the edges represent the branching structure.
}
\label{Tflip}
\end{figure}

\begin{figure}[htbp]
\begin{center}
\includegraphics[scale=0.7]{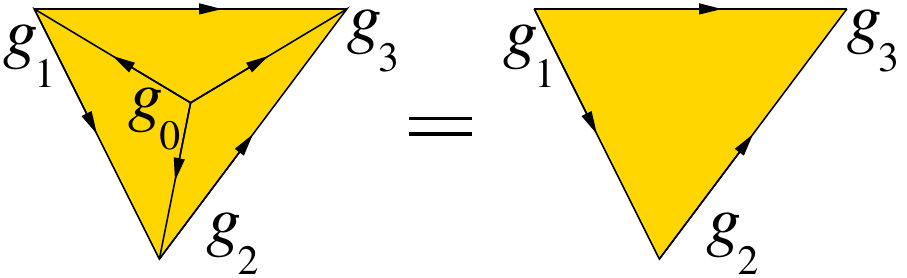}
\end{center}
\caption{
(Color online)
Graphic representation of
 $ \nu_2(g_1,g_2,g_3)= \nu_2(g_0,g_1,g_2) \nu_2(g_0,g_2,g_3) \nu_2^{-1}(g_0,g_1,g_3)$.
The arrows on the edges represent the branching structure.
}
\label{T1to3}
\end{figure}

We can also show that Eq.(\ref{Lnu2}) is a fixed-point action-amplitude from the
cocycle conditions on $\nu_2(g_i,g_j,g_k)$. The cocycle condition Eq.(\ref{condition}) 
induces two renormalization moves in the discretized manifold. 
Fig. \ref{Tflip} represents a $2\leftrightarrow2$ moves:
\begin{align}
\nu_2(g_0,g_1,g_3)\nu_2(g_1,g_2,g_3)=\nu_2(g_0,g_1,g_2)\nu_2(g_0,g_2,g_3)
\end{align}
and Fig. \ref{T1to3} represents a $1\leftrightarrow3$ move:
\begin{align}
\nu_2(g_1,g_2,g_3)=\nu_2(g_0,g_1,g_2)\nu_2(g_0,g_2,g_3)\nu_2^{-1}(g_0,g_1,g_3)
\end{align}
By using these two moves, different triangularization of the space-time can be mapped into each other without changing the action amlitude Eq.(\ref{action}). Therefore, the action-amplitude is a fixed point form.

\begin{figure}[tb]
\begin{center}
\includegraphics[scale=0.4]{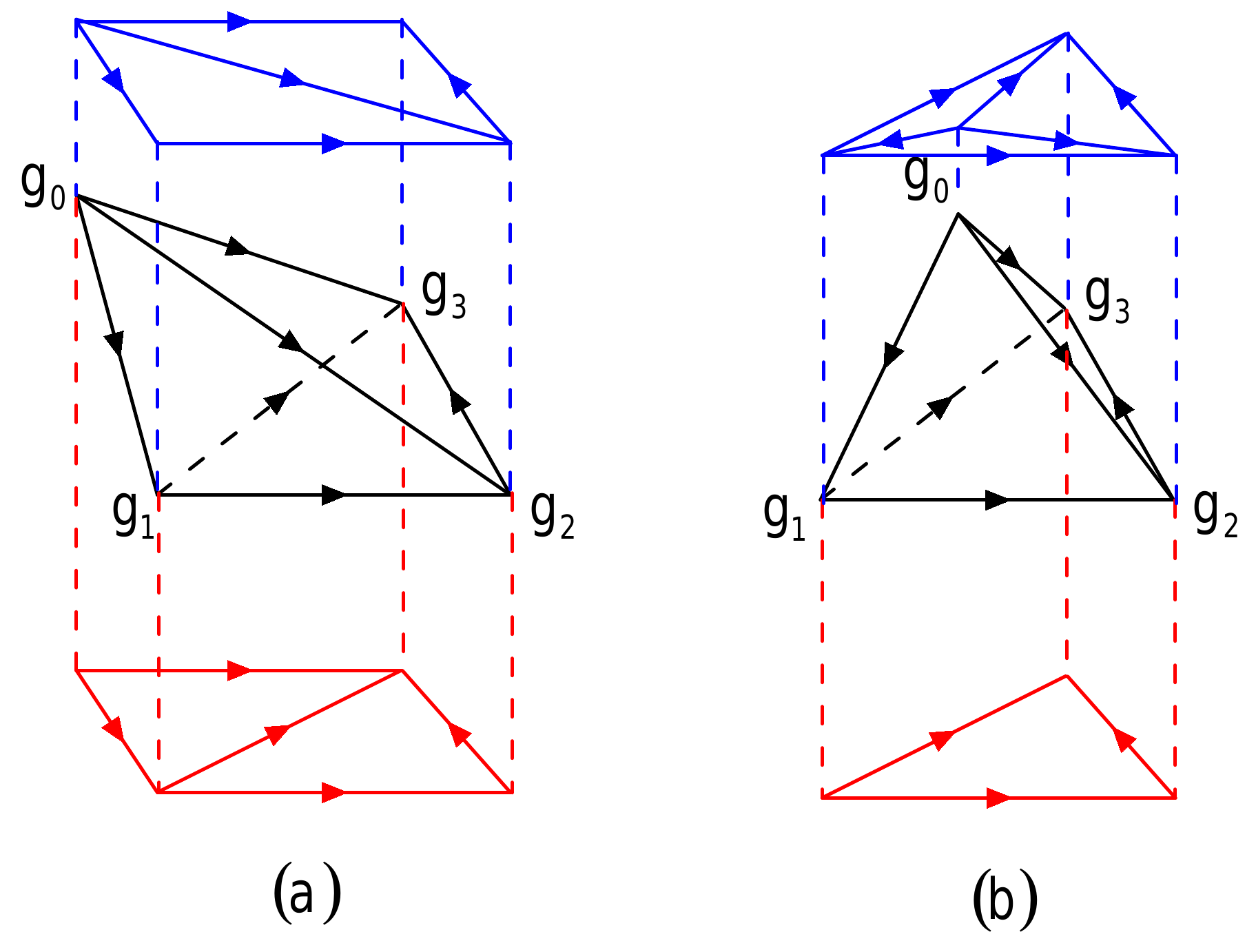}
\end{center}
\caption{$(d_2\nu_2)(g_0,g_1,g_2,g_3)$ can be represented as the
boundary of a $3$-simplex. (a) and (b) correspond to two different
basic moves of $2$-simplexes. \label{moves}}
\end{figure}

Geometrically, these two moves can be obtained by projecting a 3 simplex onto two two-dimensional surfaces. Fig.\ref{moves} (a) gives the $2 \leftrightarrow 2$ move and Fig.\ref{moves} (b) gives the $1 \leftrightarrow 3$ move. Note that the projection from opposite directions will induce
opposite chiralities for the boundary of the tetrahedron, that's
why we need to change the chiralities of the triangular in one
side of basic moves. Such a change correspond to inverse $\nu_2$
when we move it from left side to right side of the above equation,
which is consistent with our rules of algebra. Similar process also works in higher dimensions. Different moves in $d$-dimension can be obtained from projecting a $d+1$-simplex in different ways. Such moves induces a renormalization flow in the discrete space-time under which the action-amplitude Eq.(\ref{action}) is a fixed point.


\end{document}